# Bismuth-surfactant-induced growth and structure of InAs/GaAs(110) quantum dots

*Ryan B. Lewis,[1,2]\* Achim Trampert,[1] Esperanza Luna,[1] Jesús Herranz,[1] Carsten Pfüller[1] and Lutz Geelhaar[1]*

[1] Paul-Drude-Institut für Festkörperelektronik, Leibniz-Institut im Forschungsverbund Berlin e.V., Hausvogteiplatz 5-7, 10117 Berlin, Germany
[2] Department of Engineering Physics, McMaster University, L8S 4L7 Hamilton, Canada
\*Email: rlewis@mcmaster.ca

We explore the Bi-surfactant-directed self-assembly and structure of InAs quantum dots grown on GaAs(110) by molecular beam epitaxy. The addition of a Bi flux during InAs deposition changes the InAs growth mode from two-dimensional (2D) Frank–van der Merwe to Stranski–Krastanov, resulting in the formation of three-dimensional (3D) InAs islands on the surface. Furthermore, exposing static InAs 2D layers to Bi induces a rearrangement of the strained layer into 3D islands. We explore the effect of varying the InAs thickness and Bi flux for these two growth approaches, observing a critical thickness for 3D island formation in both cases. Characterization of (110) InAs quantum dots with high-resolution transmission electron microscopy reveals that larger islands grown by the Stranski–Krastanov mode are plastically relaxed, while small islands grown by the on-demand approach are coherent. Strain relaxation along the [1$\bar{1}$0] direction is achieved by 90° pure-edge dislocations with dislocation lines running along [001]. In contrast, strain relief along [001] is by 60° misfit dislocations. This behaviour is consistent with observations of planar (In,Ga)As/GaAs(110) layers. These results illustrate how surfactant Bi can provoke and control quantum dot formation where it normally does not occur.

(In,Ga)As quantum dots (QDs) are of high interest for quantum optics applications requiring single photon emission[1,2] and emission of entangled photon pairs.[3,4] While the properties of QDs are intimately linked to their symmetry—and thus the substrate orientation—InAs QD synthesis on GaAs has largely been restricted to the (100) surface, where the Stranski–Krastanov (SK) mechanism readily occurs. By contrast, the SK growth mode does not normally occur on other low-index GaAs surfaces such as (110) and (111).[5–8]

Recently, we revealed that a surface-energy-modifying Bi surfactant can induce InAs QD formation on GaAs(110) planar surfaces and nanowire sidewalls, presenting a powerful new approach for externally directing QD formation.[9,10] Such InAs/GaAs(110) QDs are expected to exhibit a low $C_s$ symmetry, and it has been predicted theoretically that strong in-plane piezoelectric fields in (110) QDs lead to linearly polarized emission associated with the ground state exciton.[11] Accordingly, recent photoluminescence (PL) experiments have indicated that emission from Bi-induced (110) QDs is linearly polarized.[12] We note that historically, surfactants have mostly been used to kinetically suppress three-dimensional (3D) island formation in strained-layer epitaxy.[13,14]

Here, we investigate the Bi-surfactant-induced self-assembly of InAs 3D islands on GaAs(110) by molecular beam epitaxy (MBE). The presence of a Bi flux during InAs deposition changes the InAs growth mode from two-dimensional (2D) layer to SK, and we explore the effect of varying the Bi flux and InAs thickness on the resulting 3D islands. In addition to the SK growth mode, we form InAs 3D islands "on-demand" by exposing static 2D InAs layers of varying thicknesses to a nominal Bi dose of 0.8 monolayers (ML). This approach is exploited to decrease the 3D island size and increase the density, providing a higher degree of control over the self-assembly process. For both SK and on-demand growth modes, we observe a critical thickness for 3D island formation. Transmission electron microscopy (TEM) reveals that the larger SK islands are plastically relaxed, while smaller islands grown by the on-demand growth mode are coherently strained.

Samples were grown by MBE on semi-insulating GaAs(110) wafers. Fluxes were provided by effusion cells for Ga, In, and Bi, and by a valved cracker for $As_2$. After thermal desorption of the native oxide, 50-nm-thick GaAs buffer layers were deposited at a substrate temperature of 420 °C and a growth rate of 0.28 ML/s [equal to 0.20 μm/h, note that 1 ML on (110) is equal to 0.7 ML on (100)]. InAs deposition was subsequently carried out with a growth rate of 0.14 ML/s. For experiments with concurrent In and Bi deposition (SK growth), the Bi flux was initiated 20 s before the In flux and maintained throughout the InAs deposition, which was carried out at a growth temperature of 420 °C. For the on-demand growth experiments, InAs layers were deposited at 370 °C in the absence of Bi, and the resulting layers were exposed to a Bi flux of 0.4 ML/s for 2 s. For all growth experiments, an $As_2$ flux of 5.7 ML/s was maintained throughout the entire deposition process. After deposition, samples were cooled at 2 °C/s while maintaining the As$_2$ flux until the substrate temperature dropped below 350 °C. Cross-sectional specimens for TEM and scanning TEM (STEM) were prepared by mechanical polishing, dimpling and precision ion-milling, which was performed with ion beam energies between 2.5 and 3 keV to minimize radiation damage. The lattice images were recorded using a Jeol 2100F microscope equipped with bright- and dark-field detectors for scanning mode. Chemically sensitive g$_{002}$ and g$_{020}$ dark-field TEM micrographs were obtained using a JEOL JEM 3010 electron microscope operated at 300 kV.

We first explore the deposition of InAs on GaAs(110) under the presence of a Bi flux. Figure 1(a-g) displays atomic force microscopy (AFM) topographs for various InAs thicknesses (increasing left–right) and Bi fluxes (increasing top–bottom). In the absence of a Bi flux [FIG. 1(a)], deposition of 2.1 ML of InAs results in a smooth 2D layer with atomic terraces visible in the topograph. The observation of 2D InAs growth on GaAs(110) in the absence of Bi is consistent with previous reports.[5–7] Figure 1(b-e) shows the surface topology for various InAs thicknesses deposited under the presence of a 0.4 ML/s Bi flux. While the surface exhibits a 2D morphology for 1.4 ML of InAs [FIG. 1(b)], a low density of 3D islands is visible on the surface for 1.7 ML [FIG. 1(c)] (including small islands of about 10 nm diameter). Further increasing the InAs thickness from 1.7 to 2.1 ML [FIGs 1(c-d)] dramatically increases the density of 3D islands from about $2\times10^8$ cm$^{-2}$ to $6\times10^9$ cm$^{-2}$. This increase in island density coincides with a characteristic 2D-to-3D transition in the reflection high-energy electron diffraction pattern (not shown). The islands in FIG. 1(d) have a height of 4.3±0.5 nm. Further increasing the InAs deposition to 5.7 ML under 0.4 ML/s Bi [FIG. 1(e)] results in an enlargement of the 3D island size and a similar island density ($5\times10^9$ cm$^{-2}$). The rapid increase and subsequent saturation in the 3D island density above a critical thickness is consistent with the SK growth mode. Thus, the Bi flux changes the InAs growth mode on GaAs(110) from 2D Frank–van der Merwe to SK. We have recently shown that this dramatic change in the surface topography is a consequence of Bi modifying surface energies.[9] The 3D islands are elongated along [1$\bar{1}$0], possibly a result of anisotropic strain relaxation,[5] or different adatom diffusivities along the in-plane [1$\bar{1}$0] and [001] directions.[15]

For the higher Bi flux of 1 ML/s and an InAs thickness of 2.1 ML [FIG. 1(f)], the 3D islands exhibit a bimodal size distribution, consisting of islands similar to those observed in FIG. 1(d) as well as larger islands with a height of 23±3 nm. Further increasing the Bi flux to 2 ML/s [FIG. 1(g)] results in a low density of even larger islands of 48±4 nm height. These islands exhibit crack-like features [highlighted in the inset of FIG 1(g)]. Energy-dispersive x-ray spectroscopy (EDS) analysis on this sample (carried out in a scanning electron microscope) reveals that the right side of these large islands are highly Bi-rich, while the left side contains (In,Ga)As and no detectable Bi. This suggests that InAs precipitates from the Bi-rich droplet (Bi is liquid at growth temperatures) and that the vapor-liquid-solid growth occurs in the [111]A direction [the right-facing facet of a 3D island is (111)A]. In contrast, EDS measurements on samples grown with lower Bi fluxes do not show the presence of Bi anywhere on the surface. However, in FIG. 1(f) depressions are visible on the right side of the larger islands (present independent of AFM scan direction). We speculate that these depressions contained Bi during the deposition, and that this Bi desorbed after the growth was interrupted. At the As-rich growth conditions used here, Bi is not expected to form a compound with InAs.[16] Furthermore, we expect that for the Bi flux of 0.4 ML/s, Bi does not accumulate on the surface but rather produces a steady-state wetting layer coverage. This assumption is supported by RHEED studies of the adsorption of Bi on GaAs(001) surfaces.[17]

The microstructure of the sample shown in FIG. 1(d) was investigated with cross-sectional TEM. Figure 1(h-i) presents high-resolution TEM (h) and STEM (i) micrographs of two separate 3D islands taken along the [$\bar{1}$10] zone axis. These micrographs illustrate that the InAs islands are primarily composed of flat (110) tops and inclined {111} sidewalls. Furthermore, both islands are semi-coherent along [001], exhibiting evidence of 60° misfit dislocations at the GaAs/InAs interface (white arrows in FIGs 1(h-i). A bright-field TEM micrograph of a 3D island taken along the orthogonal [001] zone axis is presented in FIG. 1(j). This micrograph also shows evidence of interface dislocations (white arrows). Analysis of dark-field weak-beam micrographs on similar islands taken at a projection inclined at the interface reveals that these are 90° pure-edge dislocations with Burgers vector 1/2[1$\bar{1}$0] and dislocation lines running along [001]. The observed dislocation types are consistent with previous reports of strain relaxation of InAs planar layers on GaAs(110).[5] In that study, it was also found that in planar InAs(110) layers, strain relaxation along [$\bar{1}$10] is easier. This effect could explain why our 3D islands are elongated along this direction, if 90° dislocations provide [$\bar{1}$10] strain relief before 60° dislocations relieve [001] strain.

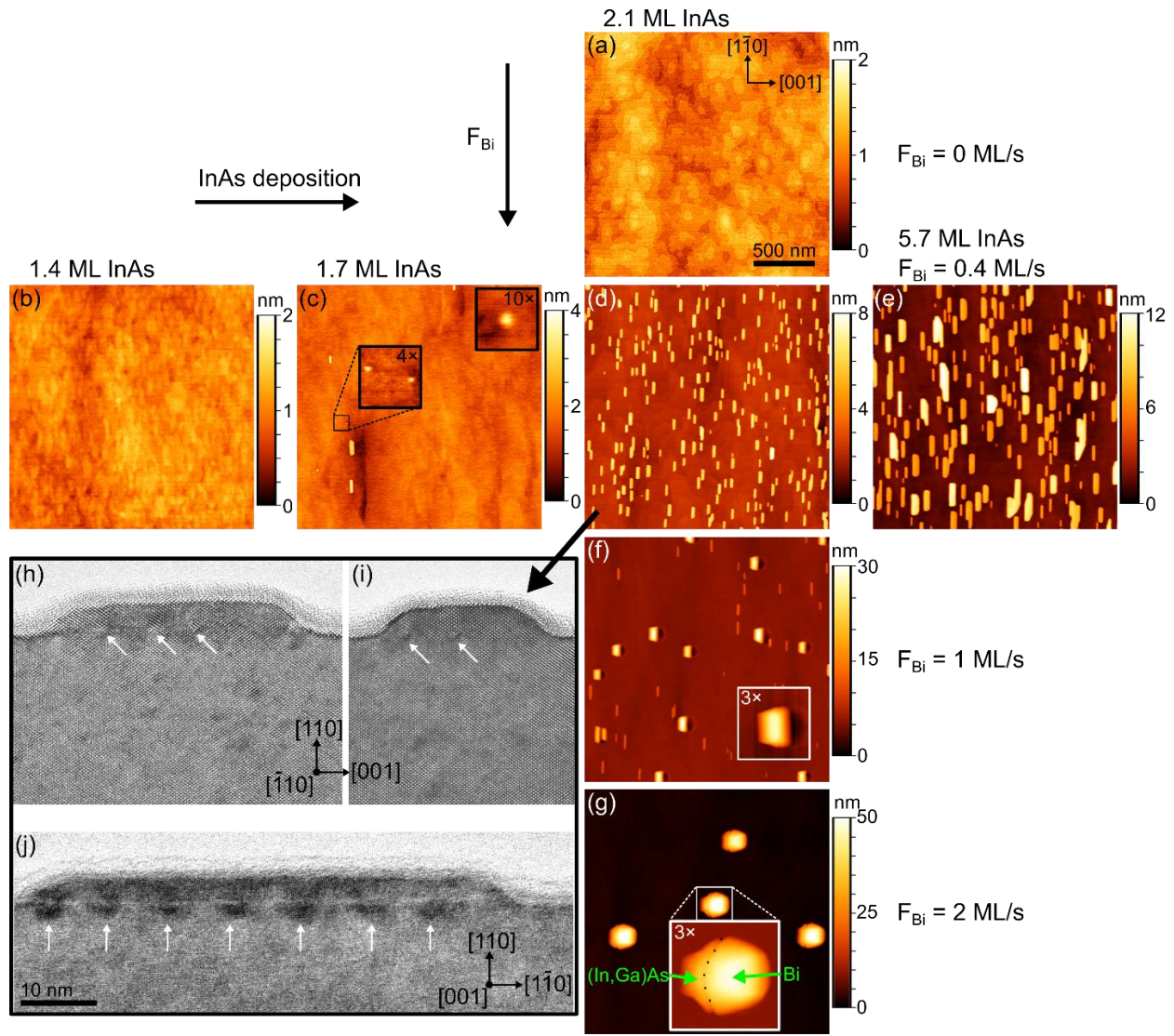

FIG. 1. (a-g) AFM topographs of InAs/GaAs(110) surfaces for InAs deposition thickness of 1.4–5.7 ML (left–right) deposited under Bi fluxes of 0–2 ML/s (top–bottom). Small 3D islands of about 10 nm diameter are visible on the surface in (c). The scale bar and crystallographic directions shown in (a) apply to all AFM topographs. (g) Only for the highest Bi flux of 2 ML/s was Bi detected on the sample by EDS. (h-j) Cross-sectional TEM micrographs of individual 3D islands from the sample shown in (d). (h-i) High-resolution TEM and bright-field STEM micrographs taken along [1̄10] showing individual islands with 60° misfit dislocations. The islands are composed primarily of {111} inclined sidewalls and (110) flat tops. (j) Bright field STEM image of a single InAs island taken along [001]. Dislocations in (h-j) are indicated by white arrows. The scale bar in (j) also applies to (h-i).

In a separate sample, 3D islands similar to those displayed in FIG. 1(h-i) were overgrown with 50 nm of GaAs immediately following the InAs deposition. Figure 2 presents chemically-sensitive {002} dark-field TEM images recorded from this sample close to (a) the [$\bar{1}$10] zone axis and (b) the orthogonal [001] zone axis. The presence of a 3–4 nm thick (In,Ga)As wetting layer is observed as a dark line in both images. The capped 3D islands (one in each image) exhibit a morphology similar to the uncapped islands.

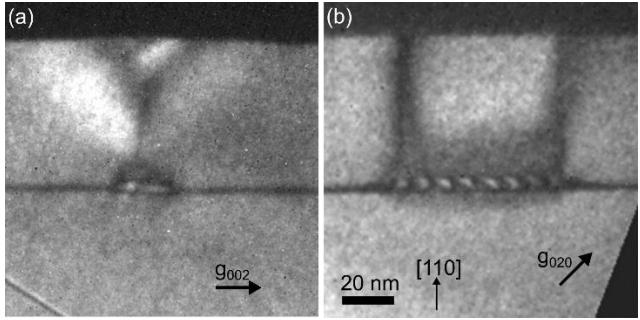

FIG 2. Chemically sensitive dark-field TEM images recorded (a) close to the [$\bar{1}$10] zone axis, and (b) close to the orthogonal [001] zone axis for a sample containing InAs islands capped with 50 nm of GaAs. The scale bar in (b) also applies to (a).

In addition to changing the growth mode of InAs on GaAs(110) from 2D Frank–van der Merwe to SK, subsequently exposing static 2D InAs layers (grown in the absence of Bi) to a Bi flux can provoke a 2D-to-3D morphological phase transition on-demand.[9] Figure 3(a-c) displays AFM topographs of surfaces where 2D InAs layers of varying thicknesses were exposed to a Bi flux of 0.4 ML/s for 2 s at 370 °C. The lower substrate temperature was selected to produce smaller InAs islands (compared to those in FIG. 1) by limiting surface diffusion. The short 2 s Bi deposition corresponds to a nominal dose of 0.8 ML, which is not expected to produce Bi surface droplets, even for low substrate temperatures where Bi desorption is slow. Smaller-diameter islands are desired as they are less likely to contain dislocations. Prior to Bi exposure, the 2D InAs layers were expected to be coherently strained to the GaAs substrate, as their thicknesses were below the critical thickness for dislocations (previously reported to be 2–3 ML).[7] While exposing a 1.3 ML InAs layer to Bi does not induce 3D island formation [FIG. 3(a)], the formation of 3D islands is observed for InAs thicknesses of 1.6–2.1 ML. This indicates an InAs critical thickness between 1.3 and 1.6 ML for this growth approach, analogous to InAs deposition in the presence of a Bi flux (SK growth). The densities of InAs islands are $7.2\times10^9$, $2.9\times10^{10}$ and $3.5\times10^{10}$ cm$^{-2}$ for InAs thicknesses of 1.6, 1.8 and 2.1 ML, respectively. For conventional SK growth of InAs on GaAs(001), varying the density over this range would require deposition thicknesses ranging from 0.02 to 0.05 ML above the SK critical thickness.[18] The extremely abrupt change in QD density near the critical thickness on (001) surfaces limits the ability to control QD densities. Density control is especially important for applications involving single QDs. One approach used to obtain low QD densities is the formation of QDs in subcritical InAs layers, which also requires very precise thickness control.[19] In comparison, for on-demand growth on (110), the density is about 27× less sensitive to the deposition thickness within this density range. Furthermore, our on-demand growth approach decouples the InAs deposition from the 3D island self-assembly process. Thus, on-demand growth could offer an unprecedented level of external control over QD synthesis. The average 3D island height is found to be independent of the initial InAs thickness (and thus the 3D island density). We note that independent control of QD density and size has previously motivated alternate growth approaches for InAs QDs, such as droplet epitaxy.[20]

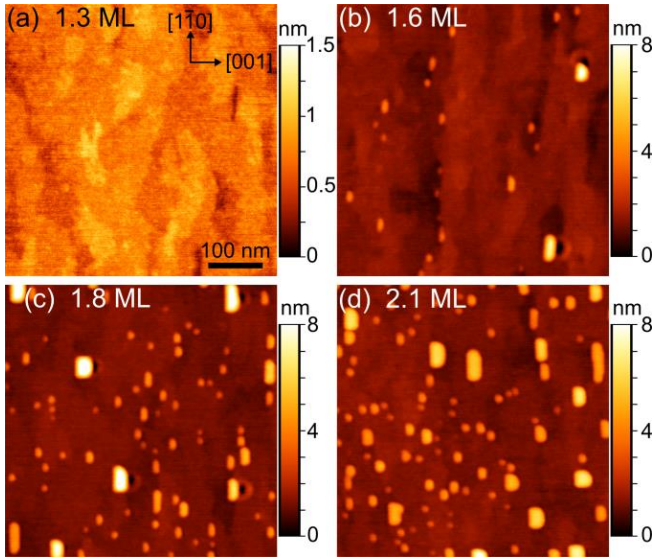

FIG. 3. AFM topographs of InAs/GaAs(110) surfaces after exposing InAs layers with various thicknesses (indicated in the figure) to a Bi flux of 0.4 ML/s for 2 s at 370 °C. (a) For 1.3 ML of InAs, the surface remains 2D and atomic terraces are visible on the surface. (b-d) For 1.6–2.1 ML of InAs, exposure of the surface to Bi induces the formation of 3D islands, the density of which increases with the initial InAs thickness, yielding $7.2 \times 10^9$, $2.9 \times 10^{10}$ and $3.5 \times 10^{10}$ cm$^{-2}$ for 1.6, 1.8 and 2.1 ML, respectively. The average height of the distributions for 1.6, 1.8 and 2.1 ML InAs thicknesses are 2.7, 3.0 and 2.7 nm, respectively.

Figure 4 shows high-resolution TEM and STEM micrographs of InAs islands from a sample grown with the same conditions as for the sample in FIG. 3(d). The images illustrate small coherent islands, with diameters of about 10 nm or less. As illustrated in FIG. 3(d), the island size varies significantly for these samples. Correspondingly, we find that small islands like those presented in FIG. 4 are coherent, while larger islands exhibit evidence of misfit dislocations (not shown). We note that the small islands in FIG. 3 have a round/symmetric shape, and thus we expect that these islands are also coherently strained along [1̄10].

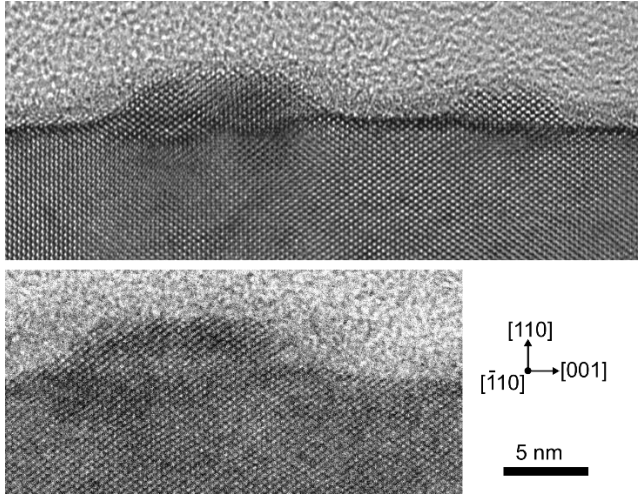

FIG. 4. [1̄10] cross-sectional high-resolution bright-field TEM (upper) and STEM (lower) micrographs of small coherent InAs 3D islands from a sample grown with the same conditions as for the sample of FIG. 3(d).

In summary, we have explored the effect of varying InAs thickness and Bi flux on the Bi-surfactant-directed growth and structure of InAs 3D islands on GaAs(110). For both InAs deposition under a Bi flux, as well as subsequently exposing static InAs layers to Bi, we observe a critical InAs thickness for 3D island formation. Structural characterization of (110) InAs islands with TEM reveals that strain relaxation along [11̄0] is achieved by 90° pure-edge dislocations, while strain relief along [001] is by 60° misfit dislocations—consistent with planar InAs films on GaAs(110). These results illustrate the great potential of surface-energy-modifying surfactants for inducing and controlling quantum dot formation in strained-layer epitaxy.

Acknowledgments

The authors are grateful to M. Höricke and C. Stemmler for MBE maintenance, S. Krauss and M. Matzeck for TEM sample preparation, and M. Niehle for a critical reading of the manuscript. R.B.L. acknowledges funding from the Alexander von Humboldt Foundation.